\documentclass[conference]{IEEEtran}
\usepackage[top=0.68in, bottom=0.68in, left=0.62in, right=0.62in]{geometry}

\usepackage{amsfonts}
\usepackage{times}
\usepackage{graphicx}
\usepackage{epstopdf}
\usepackage{latexsym}
\usepackage{dsfont}
\usepackage{amssymb}
\usepackage{amsmath}
\usepackage{cite}
\usepackage{verbatim}

\newcommand{\figref}[1]{{Fig.}~\ref{#1}}

\def\bb0{{\mathbb{0}}}

\def\ba{{\mathbf{a}}}
\def\bb{{\mathbf{b}}}

\def\bff{{\mathbf{f}}}

\def\bh{{\mathbf{h}}}

\def\bn{{\mathbf{n}}}

\def\bp{{\mathbf{p}}}

\def\bv{{\mathbf{v}}}

\def\bx{{\mathbf{x}}}
\def\by{{\mathbf{y}}}

\def\b0{{\mathbf{0}}}

\def\bA{{\mathbf{A}}}

\def\bF{{\mathbf{F}}}

\def\bH{{\mathbf{H}}}
\def\bI{{\mathbf{I}}}

\def\bQ{{\mathbf{Q}}}

\def\bbC{{\mathbb{C}}}

\def\bbE{{\mathbb{E}}}

\def\bbS{{\mathbb{S}}}

\def\cC{\mathcal{C}}

\def\cE{\mathcal{E}}

\def\cN{\mathcal{N}}

\def\sf0{{\mathsf{0}}}

\usepackage{graphicx}
\usepackage{epstopdf}
\usepackage{enumerate}
\usepackage{algorithm}
\usepackage{amsmath}
\usepackage{mathrsfs}
\usepackage{float}
\usepackage{color}
\usepackage{makeidx}
\usepackage{bm}
\usepackage{cleveref}
\usepackage{url}
\usepackage{steinmetz}
\usepackage{varwidth}
\usepackage{soul}
\usepackage{bbm}
\usepackage{empheq}
\usepackage{mathtools}
\usepackage{cite}
\usepackage{balance}
\usepackage{subfigure}

\newcommand{\sref}[1]{{Section}~\ref{#1}}

\usepackage{amsfonts}
\usepackage{booktabs}
\usepackage{siunitx}

\begin{document}
	\title{Digital Twin Assisted Beamforming Design for Integrated Sensing and Communication Systems}
	\author{Shuaifeng Jiang and Ahmed Alkhateeb \\School of Electrical, Computer, and Energy Engineering, Arizona State University\\Emails: \{s.jiang, alkhateeb\}@asu.edu}

\maketitle
\begin{abstract}
	This paper explores a novel research direction where a digital twin is leveraged to assist the beamforming design for an integrated sensing and communication (ISAC) system. In this setup, a base station designs joint communication and sensing beamforming to serve the communication user and detect the sensing target concurrently. Utilizing the electromagnetic (EM) 3D model of the environment and ray tracing, the digital twin can provide various information, e.g., propagation path parameters and wireless channels, to aid communication and sensing systems. More specifically, our digital twin-based beamforming design first leverages the environment EM 3D model and ray tracing to (i) predict the directions of the line-of-sight (LoS) and non-line-of-sight (NLoS) sensing channel paths and (ii) identify the dominant one among these sensing channel paths. Then, to optimize the joint sensing and communication beam, we maximize the sensing signal-to-noise ratio (SNR) on the dominant sensing channel component while satisfying a minimum communication signal-to-interference-plus-noise ratio (SINR) requirement. Simulation results show that the proposed digital twin-assisted beamforming design achieves near-optimal target sensing SNR in both LoS and NLoS dominant areas, while ensuring the required SINR for the communication user. This highlights the potential of leveraging digital twins to assist ISAC systems.
\end{abstract}


\section{Introduction}\label{Introduction}
Integrated sensing and communication (ISAC) systems have emerged as a pivotal technology in the evolution of wireless communications \cite{liu2022integrated,demirhan2023integrated}. By combining the functionalities of communication and sensing into a unified framework, ISAC is envisioned to enhance spectrum efficiency, lower hardware costs, and reduce power consumption. Moreover, the dual capability of ISAC can enable advanced applications such as autonomous driving, smart cities, and intelligent manufacturing, where real-time environment awareness and reliable communication are essential. One of the key challenges in realizing the vision of ISAC systems lies in the efficient allocation of resources for both communication and sensing tasks. To that end, in this paper, we investigate
the joint sensing and communication beamforming design of the MIMO ISAC systems.

Prior work has investigated the joint beamforming design for ISAC systems \cite{liu2020joint, luo2020multibeam, zhao2022joint, liyanaarachchi2023joint, demirhan2023cell, hua2023optimal}. \cite{liu2020joint} proposed joint transmit beamforming approaches for co-located MIMO radar and multiuser MIMO communication systems. In \cite{luo2020multibeam} the author proposed a global optimization method that directly optimizes the joint sensing and communication beamforming vector for analog antenna arrays. \cite{zhao2022joint} studies the joint transmit and receive ISAC beamforming considering the imperfect channel information. In \cite{liyanaarachchi2023joint} the author developed the joint hybrid beamforming and waveform optimization for ISAC systems. \cite{demirhan2023cell} considered cell-free MIMO ISAC systems, and derived the joint sensing and communication beamforming design based on a max-min fairness communication criterion. In \cite{hua2023optimal}, the authors investigated the ISAC beamforming design in the case where the communication users have the capability of canceling the interference from predefined sensing signals. While these studies have shown promising results, they are predominantly limited by (one of the) two key assumptions. (i) The sensing target is line-of-sight (LoS) to the BS, allowing the ISAC system to design the sensing beam along the LoS direction. (ii) The ISAC system has some knowledge of the wireless sensing channel between the BS and the target. The first assumption hinders the ISAC system from detecting non-line-of-sight (NLoS) targets. The second assumption poses challenges in obtaining the sensing channel information: The sensing target often cannot cooperate in pilot-based channel estimation, and the sensing channel is related to the shape and material of the unknown target itself.

Digital twin-aided communication has recently gained increasing interest \cite{alkhateeb2023real, jiang2024learnable}. The digital twin approximates the real-world communication environment with the electromagnetic (EM) 3D model. By applying accurate ray tracing on the EM 3D model, the digital twin can infer/predict information about the real-world wireless channel with even eliminated channel acquisition overhead. The predicted channel information can be used to develop wireless datasets to train machine learning models \cite{alikhani2024large}, or can be used to directly aid real-time communication operations. This capability is interesting for sensing or ISAC applications where obtaining the sensing channel is challenging. Based on that, his paper introduces a novel digital twin-assisted beamforming approach for ISAC systems. In particular, the digital twin is leveraged to predict the directions and partial channel gains of both LoS and NLoS sensing channel paths. This sensing channel information is then used to guide the design of the efficient ISAC beamforming. The contribution of this paper is threefold.
\begin{itemize}
	\item We proposed to leverage digital twins to infer information about the wireless sensing channel, thereby assisting the operation of ISAC systems.
	\item We first develop a joint communication and sensing beamforming design that leverages the dominant NLoS direction of the sensing target predicted by the digital twin, which achieves high sensing performance for NLoS sensing directions. 
	\item We then propose a joint communication and sensing beamforming design that utilizes the partial channel information predicted by the digital twin, which can adapt to both LoS and NLoS sensing directions. 
\end{itemize}

We evaluate the proposed digital twin-aided ISAC beamforming on wireless data generated from high-fidelity ray tracing. The simulation results are compelling, demonstrating that our joint beamforming approach can achieve near-optimal target sensing sensing signal-to-noise ratio (SNR) while ensuring the required signal-to-interference-plus-noise ratio (SINR) for the communication user. Importantly, our approach does not assume any information about the sensing channel except for the candidate sensing target position. This underscores the potential of leveraging digital twins to assist ISAC systems.

\section{System Model}\label{sec:system_model}
We consider a MIMO ISAC system incorporating one base station (BS), one communication user equipment (UE), and one sensing target. In the considered system, the BS transmits a joint communication and sensing (JSAC) signal to serve the user and simultaneously detect the sensing target. The BS is equipped with $N_t$ transmitting antennas and $N_r$ receiving antennas with full digital beamforming capability, \textit{i.e.}, each antenna element has a dedicated radio frequency (RF) chain. For simplicity, we assume that the BS can perfectly cancel the self-interference, \textit{i.e.}, and separate the received signal from the transmitted signal without signal leakage. The UE is equipped with a single antenna.

Moreover, we assume that the BS has access to a static digital twin of its environment. The static digital twin incorporates the 3D EM models of all the static objects. Given the position and orientation of a transmitter and a receiver, the digital twin can apply accurate ray tracing to the 3D EM models to synthesize the channel between them. The BS leverages this digital twin to achieve the ISAC task.

For the communication sub-task, we consider the downlink data transmission, \textit{i.e.}, the BS transmits the JSAC signal to send downlink data to the UE. For the sensing sub-task, we consider the target detection task, \textit{i.e.}. In particular, given a target position, the BS transmits the JSAC signal, receives the JSAC signal distorted by the environment, and determines whether a target is present at the position or not.

\begin{figure}[t]
	\centering
	\includegraphics[width=0.9\linewidth]{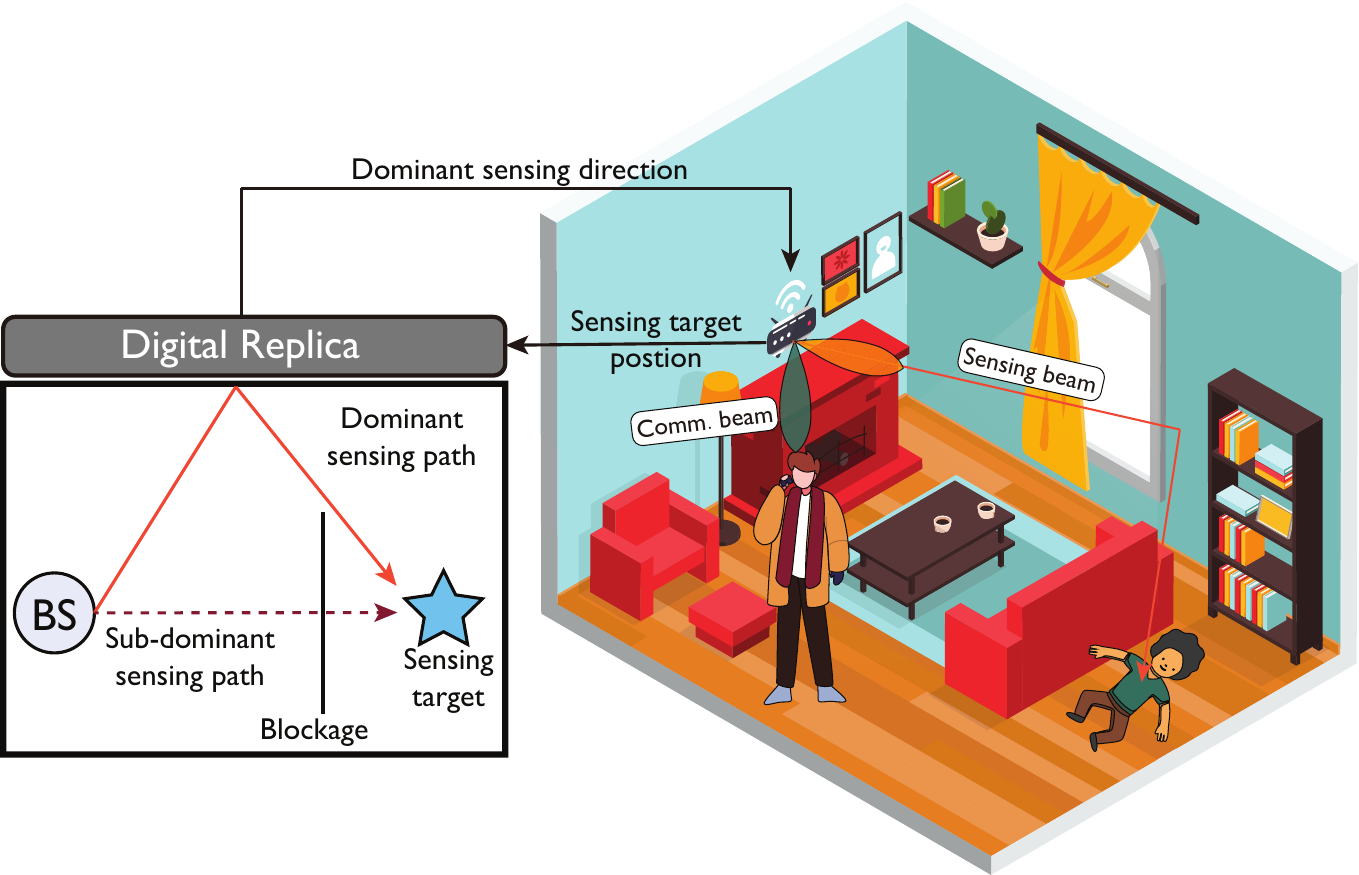}
	\caption{This figure presents the key idea of leveraging the digital twin to design the joint communication and sensing beamforming. The digital twin can provide partial channel information to guide the beamforming.}
	\label{fig:key_ida}
\end{figure}

\subsection{Signal Model}
The JSAC signal transmitted by the BS is defined as the sum of the communication and sensing signals. Let $s_u,s_t\in\bbC$ denote the communication and sensing symbols, respectively. The JSAC signal $\bx\in\bbC^{N_t\times 1}$ can then be written as
\begin{align}
	\bx = \bff_u s_u + \bff_t s_t,
\end{align}
where $\bff_u, \bff_t \in\bbC^{N_t\times 1}$ are the transmit beamforming vectors for the communication and sensing symbols, respectively. We assume that the communication and sensing symbols have unit power, \textit{i.e.}, $\bbE[s_u^Hs_u]=[s_t^Hs_t]=1$. Moreover, the communication and sensing symbols are statistically independent, \textit{i.e.}, $\bbE[s_u^Hs_t]=0$. The JSAC signal satisfies the transmit power constraint:
\begin{align}
	\bbE[x^Hx] = \left\|\mathbf{f}_t\right\|^2+\left\|\mathbf{f}_u\right\|^2 \leq P,
\end{align}
where $P$ is the total transmit power.

\subsection{Communication Model}
We denote the communication channel between the BS and the UE as $\bh_u\in\bbC^{N_t \times 1}$. The received signal at the UE can then be written as
\begin{align}\label{eq:ue_receive}
	y_u= & \bh_u^H\bx + n_u \\
	= & \bh_u^H \bff_u s_u + \bh_u^H \bff_t s_t + n_u,
\end{align}
where $n_u\sim\cC\cN(0, \sigma_u^2)$ is the additive Gaussian noise at the UE. From \eqref{eq:ue_receive}, we can write the UE's receive SINR as
\begin{align}\label{eq:comm_sinr}
	\mathrm{SINR}_u = \frac{|\bh^H_u\bff_u|^2}{|\bh^H_u\bff_t|^2 + \sigma_u^2}
\end{align}

\subsection{Sensing Model.}
We consider a general multi-path far-field geometry channel model for the sensing channel. In particular, the signal transmitted by the BS can first interact with the environment and then reach the sensing target. After that, the sensing target scatters the signal, and the scattered signal can again interact with the environment before reaching the receive antennas at the BS. For the sensing target, we model it as a single-point reflector, as commonly adopted in the literature. Following this model, the sensing channel $\bH_t\in\bbC^{N_t \times N_r}$ is given by
\begin{align}\label{eq:radar_channel}
	\bH_t = \sum_{l=1}^L\alpha_l\ba(\phi^\mathrm{AoA}_l, \theta^\mathrm{AoA}_l)\ba^T(\phi^\mathrm{AoD}_l, \theta^\mathrm{AoD}_l),
\end{align}
where $L$ denotes the total number of paths. $\phi^\mathrm{AoD}_l$ and $\theta^\mathrm{AoD}_l$ denote the azimuth and elevation angles of departure (AoD), respectively. $\phi^\mathrm{AoD}_l$ and $\theta^\mathrm{AoD}_l$ denote the azimuth and elevation angles of departure (AoD). $\ba(\phi, \theta)$ denotes the array response vector of the transmit or receive antenna arrays given. $\alpha_l$ is the complex path gain of the $l$-th path, which can be modeled as
\begin{align}\label{eq:decomp}
	\alpha_l &=  \alpha_{l,j_l} \prod_{i=1}^{j_l-1} \alpha_{l,i} e^{-j2\pi f_c\tau_{l,i}} \prod_{i=j_l+1}^{I_l} \alpha_{l,i} e^{-j2\pi f_c\tau_{l,i}}\\
	&= \alpha_{l,j_l} \beta_{l,1}\beta_{l,2},
\end{align}
where $\tau_{l,i}$ and $\alpha_{l,i}$ are the propagation delay and complex gain of the $i$-th interaction ($i=1,\hdots,I_l$) along the $l$-th path. $f_c$ is the carrier frequency. At the $j_l$-th interaction, the sensing target distorts the signal. $\beta_{l,1}$ and $\beta_{l,2}$ represent the total complex gain of the propagation before and after the sensing target, respectively. 

With \eqref{eq:radar_channel}, the received JSAC signal at the BS $\by_t\in\bbC^{N_r\times 1}$ can be written as
\begin{align}\label{eq:bs_receive}
	\by_t= & \bH_t\bx + \bn_t \\
	= & \bH_t \bff_u s_u + \bH_t \bff_t s_t+ \bn_t,
\end{align}
where $\bn_t\in\bbC^{N_r \times 1}$ is the additive noise at the BS's receive antennas, which follows $\bn_t\sim\cC\cN(0, \sigma_t^2\bI)$. From \eqref{eq:bs_receive}, the sensing SNR can be written as
\begin{align}\label{eq:sensing_snr}
	\mathrm{SNR}_t = \frac{\|\bH^H_t\bff_u\|^2 + \|\bH^H_t\bff_t\|^2}{\sigma_t^2}
\end{align}
Comparing \eqref{eq:comm_sinr}. and \eqref{eq:sensing_snr}, we make an important observation on the difference between communication and sensing tasks in the JSAC system. For a communication task, the sensing signal acts as \textit{interference}. For a sensing task, however, the communication signal can \textit{contribute positively}.

\section{Problem Formulation}
In this work, we consider the downlink data transmission for the communication task and the target detection for the sensing task. The target detection task is defined as maximizing the sensing SNR given a sensing target position, which is critical for downstream sensing tasks such as target scanning and tracking. Our objective is then to design the communication beam $\bff_u$ and the sensing beam $\bff_t$ to optimize the communication SINR and the sensing SNR defined in \eqref{eq:comm_sinr}. and \eqref{eq:sensing_snr}. In particular, we aim to maximize the sensing SNR while satisfying a minimum communication SINR $\gamma_u$ to support communication applications. We provide the mathematical problem formulation as follows.
\begin{subequations}\label{eq:opt1}
	\begin{align}
		\max _{\bff_u, \bff_t} &\ \mathrm{SNR}_t \\
		\text { s.t. } & \operatorname{SINR}_u \geq \gamma_u \label{eq:12b}\\
		& \left\|\bff_t\right\|^2+\left\|\bff_u\right\|^2 \leq P, \label{eq:12c}
	\end{align}
\end{subequations}

The optimal design of the communication and sensing beams requires information about the communication channel $\bh_u$ and sensing channel $\bH_t$. The BS can obtain the communication channel by pilot-based channel estimation. Moreover, the BS may also use the digital twin to aid the channel estimation or even directly infer the communication channel. Therefore, we can assume that the communication channel is known at the BS for simplicity. In the next section, we introduce the proposed ISAC beamforming approaches.
\section{Baseline Solutions}
In this section, we present two baseline ISAC approaches.

\subsection{ISAC with Full Sensing Channel}\label{sec:full}
The optimization problem in \eqref{eq:opt1} requires knowledge about the communication and sensing channels. The communication channel can be obtained with channel estimation. For the sensing channel, we first assume it is known and try to solve the ISAC problem. Note that the sensing channel is difficult to obtain in real-world applications. Therefore, this ISAC approach utilizing the full sensing channel can be considered as an unachievable upper bound.

The maximization of the convex expression in \eqref{eq:opt1} is non-convex. To that end, we cast it into a convex problem using semidefinite relaxation \cite{luo2010semidefinite}. We re-define the optimization variables $\bff_t$ and $\bff_u$ as matrices $\bF_t=\bff_t\bff_t^H$ and $\bF_u=\bff_u\bff_u^H$. The optimization objective in \eqref{eq:opt1} becomes
\begin{subequations}
	\begin{align}
		\max_{\bF_u, \bF_t} &\ \mathrm{trace}(\bQ_t\bF_t) + \mathrm{trace}(\bQ_t\bF_u) \\
		\text { s.t. } & \mathrm{rank}(\bF_t)=1, \mathrm{rank}(\bF_u)=1\\
		& \bF_t\in\bbS^+, \bF_u\in\bbS^+,
	\end{align}
\end{subequations}
where $\bQ_t = \bH_t^H\bH_t$ and $\bbS^+$ denotes the set of hermitian positive semidefinite matrices. Note that two new constraints are introduced due to the structure of $\bF_u$ and $\bF_t$. Next, we represent the constraints in \eqref{eq:12b} and \eqref{eq:12c} with the new optimization variables $\bF_u$ and $\bF_t$. The communication SINR constraint in \eqref{eq:12b} can be re-written as follows
\begin{align}
	\frac{1}{\gamma_u}\mathrm{trace}({\bQ_u\bF_u}) - \mathrm{trace}({\bQ_u\bF_t}) \geq \sigma_u^2,
\end{align}
where $\bQ_u = \bh\bh^H$. The total transmit power constraint in \eqref{eq:12c} can be re-written as
\begin{align}
	\mathrm{trace}({\bF_u})+ \mathrm{trace}({\bF_t}) \leq 1.
\end{align}
We re-state the optimization problem using the new optimization variable $\bF_t$ and $\bF_t$
\begin{subequations}
	\begin{align}\label{eq:opt_final}
		\max_{\bF_u, \bF_t} &\ \mathrm{trace}(\bQ_{t}\bF_t) + \mathrm{trace}(\bQ_{t}\bF_u) \\
		\text { s.t. } & \mathrm{rank}(\bF_t)=1, \mathrm{rank}(\bF_u)=1 \label{eq:rank1}\\
		& \bF_t\in\bbS^+, \bF_u\in\bbS^+\\
		&  \mathrm{trace}({\bF_u}) +\mathrm{trace}({\bF_t}) \leq 1\\
		& \frac{1}{\gamma_u}\mathrm{trace}({\bQ_u\bF_u}) -\mathrm{trace}({\bQ_u\bF_t}) \geq  \sigma_u^2.
	\end{align}
\end{subequations}
This optimization problem can be relaxed by removing the rank-1 constraint in \eqref{eq:rank1} to be solved via convex optimization solvers. Note that the resulting solution for $\bF_t$ and $\bF_u$ may not be rank-1. We use the SVD decomposition to construct the rank-1 approximation of $\bF_t$ and $\bF_u$. $\bff_t$ and $\bff_u$ can then be obtained by
\begin{subequations}
	\begin{align}
		\bff_t = \sqrt{\lambda_t} \bv_t\\
		\bff_u = \sqrt{\lambda_y} \bv_y,
	\end{align}
\end{subequations}
where $\lambda_t$ is the largest singular value of $\bF_t$ and $\bv_t$ is the corresponding singular vector. $\lambda_u$ and $\bv_u$ are also similarly defined by $\bF_u$.

\subsection{ISAC with LoS Sensing Direction}\label{sec:los}
Obtaining the full sensing channel is difficult in real-world applications for the following reasons. (i) The sensing target does not have antennas and, thus, cannot cooperate in pilot-based channel estimation. (ii) Estimating the sensing channel and detecting the sensing target forms a chicken-egg problem. To that end, we propose a baseline method that does not rely on the full sensing channel information.

Let $\bp_s = [x_s, y_s, z_s]$ and $\bp_b = [x_b, y_b, z_b]$ denote the known (candidate) sensing target and BS positions, respectively. The propagation angle of the direct backscattering sensing path can be obtained by
\begin{align}
	\theta^\mathrm{AoD}_{LoS} &= \theta^\mathrm{AoA}_{LoS} = \arctan\bigg(\frac{z_s-z_b}{\sqrt{(x_s-x_b)^2 + (y_s-y_b)^2}}\bigg)\\
	\phi^\mathrm{AoD}_{LoS} &= \phi^\mathrm{AoD}_{LoS} = \arccos\bigg(\frac{y_s-y_b}{x_s-x_b}\bigg).
\end{align}
With the LoS backscattering angles, the baseline approach aims to optimize the sensing SNR of the LoS backscattering path in the sensing channel $\bH_t$. The sensing channel of the LoS backscattering path is given by
\begin{align}
	\bH_{LoS} &= \alpha_{LoS} \ba(\phi^\mathrm{AoA}_{LoS}, \theta^\mathrm{AoA}_{LoS})\ba^T(\phi^\mathrm{AoD}_{LoS}, \theta^\mathrm{AoD}_{LoS})\nonumber \\
	& = \alpha_{LoS}\bA_{LoS}.
\end{align}
The sensing and communication beams can be solved by substituting $\bQ_t$ in \eqref{eq:opt_final} with
\begin{align}\label{eq:qt_los}
	\bQ_t&=\bA_{LoS}^H\bA_{LoS}\\
	&=\ba^*(\phi^\mathrm{AoD}_{LoS})\ba^T(\phi^\mathrm{AoD}_{LoS}),
\end{align}
where $\ba^*$ denote the conjugate of $\ba$. Note that the unknown path gain $\alpha_{LoS}$ in $\bH_{LoS}$ does not affect the optimization solution in \eqref{eq:opt_final}.

\section{Digital Twin-Aided ISAC: Sensing Through the NLoS Direction}
In the previous section, we proposed two baseline ISAC beamforming approaches. (i) The first baseline approach can achieve optimal performance with the full sensing channel information. However, obtaining the full sensing channel information is challenging in real-world deployments. (ii) With (only) the LoS direction of the sensing target, we expect the second baseline approach to achieve near-optimal when the sensing channel is dominated by the LoS component. However, its performance may degrade significantly in complex real-world scenarios with multiple blockages, reflectors, and scatters. In this section, we propose to use digital twin \cite{alkhateeb2023real} to obtain prior information about the sensing channel and aid the ISAC beamforming.

\subsection{Key Idea: Digital Twin-Aided ISAC}
The real-world wireless (both communication and sensing) channel is primarily determined by the wireless environment denoted by $\cE$ and the wireless signal propagation law denoted by $g(\cdot)$. The wireless environment $\cE$ contains information about the positions, orientations, dynamics, shapes, and materials of the BS, the UE, and other objects (reflectors/scatterers). The wireless signal propagation law $g(\cdot)$ governs how EM waves are distorted by the propagation media and other objects along its propagation paths. The real-world channel $\bH$ can be written as
\begin{align}
	\bH = g(\cE).
\end{align}

Digital twins have been proposed to obtain various types of prior information, including the channel, to aid communications \cite{alkhateeb2023real}. In particular, digital twins approximate the communication environment $\cE$ with EM 3D models and the signal propagation law with ray tracing. The EM 3D model, denoted by $\widetilde{\cE}$, contains the estimated real-time geometry information and EM material information about the BS, the UE, and other static or dynamic objects (reflectors/scatterers) in the communication environment. This EM 3D model can be constructed from real-time multi-modal sensing and/or offline database. The ray tracing, denoted by $\widetilde{g}(\cdot)$, tracks the propagation path between the transmit and receive antennas based on the 3D geometry model of the communication environment. For each path, the ray tracing simulates the path parameters, including the propagation delay and angles from the geometry information and the path gain from both the geometry and material information. With the EM 3D model and the ray tracing, the digital twin can approximate the real-world channel using
\begin{align}\label{eq:dt}
	\widetilde{\bH} = \widetilde{g}(\widetilde{\cE}).
\end{align}

Ideally, the digital twin contains real-time information about the communication environment. Constructing such a real-time digital twin requires updating the information about the dynamic objects in real-time. However, this may not always be feasible due to various considerations such as cost, power consumption, and computation complexity. In this work, we consider a relatively low-cost digital twin that only contains information about the static objects in the environment and leverage that to aid the ISAC task.

\subsection{Sensing Though the Direction of the Strong Partial Path}\label{sec:nlos_approach}
The considered static digital twin cannot directly infer the sensing channel about the unknown sensing target because it is not captured in the static EM 3D model. In particular, the digital twin cannot infer the precise complex gain of the sensing path because the complex gain $\alpha_{l,j_l}$ depends on the EM property of the unknown sensing target. Moreover, how the sensing target distorted the incoming wireless signal also affects the propagation of the distorted signal afterward. For example, depending on the shape of the sensing target, the outgoing distorted signal can propagate in different directions. That is, with the unknown sensing target, $\beta_{l,2}$ is also unknown.

While the digital twin cannot directly provide the sensing channel information, it provides some prior information that can be leveraged to aid the ISAC beamforming. (i) The digital twin contains the geometry information in the 3D map, which can be used to infer the direction of all LoS and NLoS paths. (ii) With the geometry and material information, the digital twin can predict the partial path gain $\beta_{l,1}$ until the sensing path reaches the sensing target. Intuitively, a sensing path with a higher partial path gain $\beta_{l,1}$ is more likely to exhibit a greater total path gain. This assumption is reasonable, especially when the shape and material information about the sensing target is not available. Following this idea, we propose to use the digital twin to apply (partial) ray tracing between the BS and the sensing target (position) to find all propagation paths. For each path, the digital twin provides $(\beta_{l,1}, \phi^{AoD}_{l} ,\theta^{AoD}_{l})$. Then, we find the path with the highest partial gain $\beta_{\star,1}$. Let $\phi^{AoD}_{\star} ,\theta^{AoD}_{\star}$ denote the corresponding departure directions. We aim to optimize the sensing SNR of the strong partial path direction $(\phi^{AoD}_{\star} ,\theta^{AoD}_{\star})$ while ensuring a minimum communication SINR. To that end, we substitute $\phi^{AoD}_{\star}$ and $\theta^{AoD}_{\star}$ into \eqref{eq:qt_los} to obtain $\bQ_t$, and solve the optimization problem in \eqref{eq:opt_final}.

\section{Simulation Results}
In this section, we first introduce our scenario setup and data generation process. After that, we evaluate the proposed digital twin-aided ISAC performance.
\subsection{Scenario Setup}
\begin{figure}[t]
	\centering
	\includegraphics[width=0.9\linewidth]{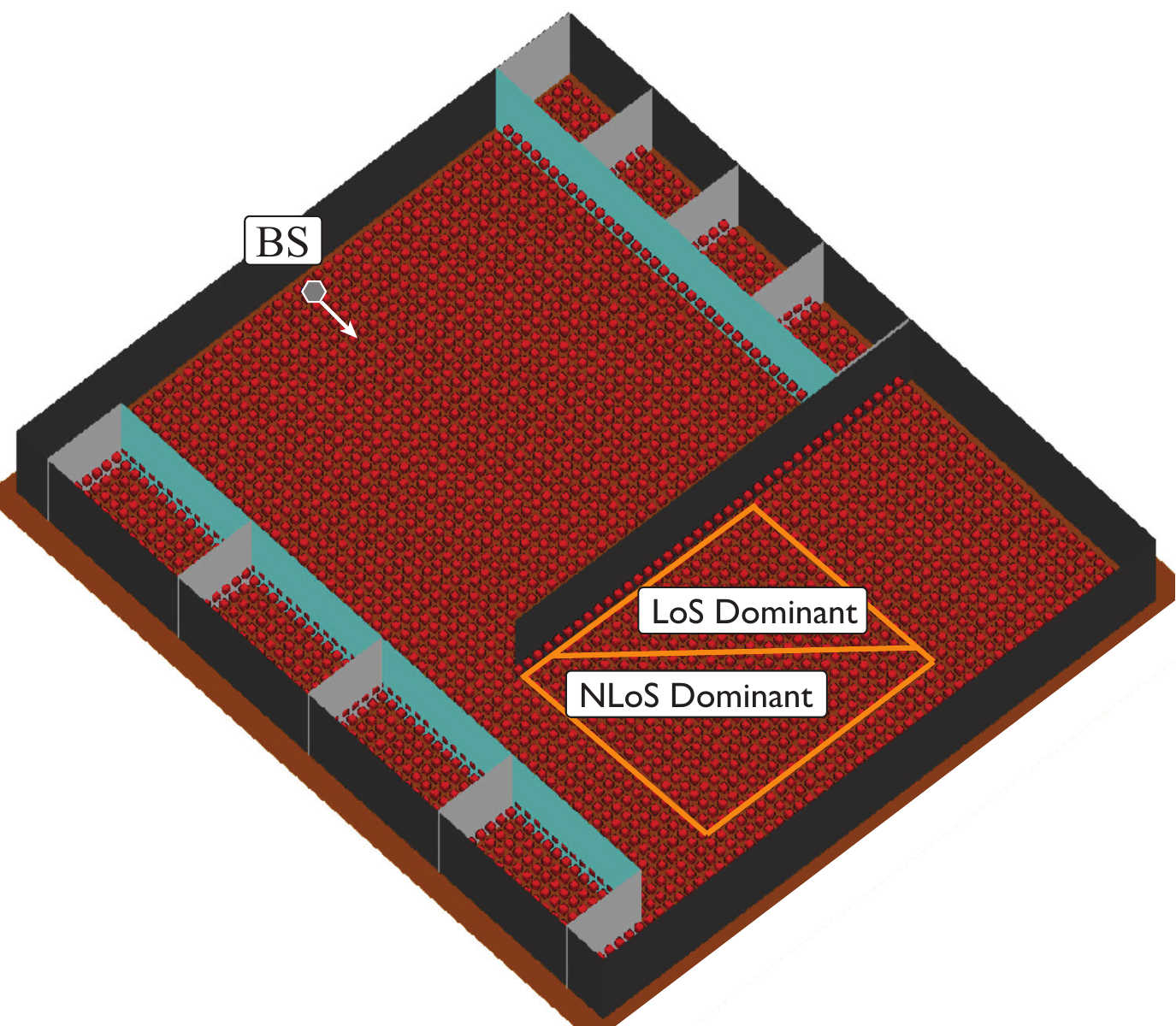}
	\caption{This figure shows the geometry layout of the adopted indoor scenario. The sensing target is placed in the area annotated in orange. The communication user grid (annotated in red) covers the entire indoor space.}
	\label{fig:scenario}
\end{figure}

We construct an indoor scenario as depicted in \figref{fig:scenario}, encompassing a room of 30 by 40 meters. The BS station and the sensing targets are positioned at two opposite sides of the room. For the sensing target, we adopt a spherical object with a diameter of 1 meter. The sensing target is placed in the area annotated in orange, and is blocked from the BS by a concrete wall of $0.2$ meters thick. The communication users are uniformly distributed in a grid that covers the entire area. The BS is equipped with $16$ transmit antennas and $16$ receive antennas. The user employs one receive antenna. The system operates at a carrier frequency of $3.5$ GHz.
\subsection{Data Generation}
To generate the communication channel data, we apply ray tracing between the BS and each user in the user grid to get the path parameters, including the path gain, propagation angle, and propagation delay. After that, we use the DeepMIMO data generator to generate the channel between the BS and each user. The communication dataset has 3233 user channels, and each user channel can be written as $\bh\in\bbC^{16\times 1}$.

To generate the sensing data, we randomly place the sensing target at 1000 positions in the sensing target area (annotated in orange) and apply ray tracing between the transmit and receive antennas of the BS. From the traced paths, we extract the paths that interact with the sensing target and generate the sensing channels. From the position of the sensing target, we obtain the LoS direction $(\phi^{AoD}_{LoS} ,\theta^{AoD}_{LoS})$. We also place a receiver antenna at the sensing target position to conduct the partial ray tracing and generate the path parameters $(\beta_{l,1}, \phi^{AoD}_{l} ,\theta^{AoD}_{l})$, and extract $(\phi^{AoD}_{\star} ,\theta^{AoD}_{\star})$ for the dominant direction. The sensing dataset consists of 1000 data samples. Each data sample has the sensing channel $\bH\in\bbC^{16\times 16}$, the sensing target position $\bp_s$, the LoS direction $(\phi^{AoD}_{LoS} ,\theta^{AoD}_{LoS})$, and the dominant direction $(\phi^{AoD}_{\star} ,\theta^{AoD}_{\star})$. In each repeated experiment, we randomly sample one communication channel from the 3233 users and one sensing channel from the 1000 sensing positions.

\subsection{Evaluation Results}
\begin{figure}[t]
	\centering
	\includegraphics[width=1.\linewidth]{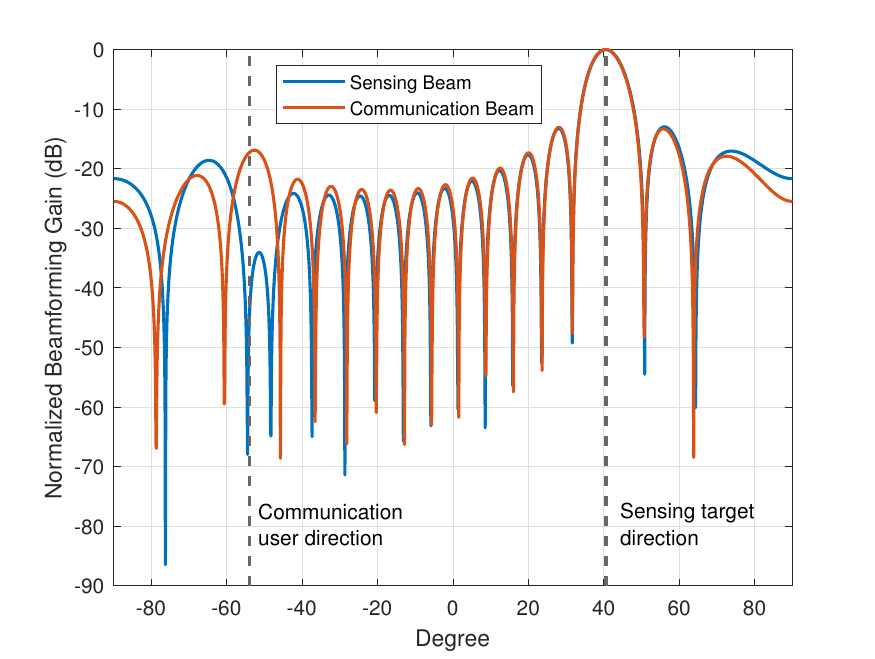}
	\caption{This figure presents the beam patterns of the sensing and the communication beams obtained by the proposed digital twin approach. The main lobes of sensing and communication beams point toward the sensing target direction. The sensing beam forms a null at the direction of the communication user to alleviate interference.}
	\label{fig:beam_pattern}
\end{figure}

In \figref{fig:beam_pattern}, we present the beam patterns of the proposed digital twin aided beamforming approach. It can be seen that the main lobes of the communication beams point toward the directions of the communication user and the sensing target, respectively. From that, the power of the communication signals is re-used for sensing. In contrast, the sensing beam forms a main lobe towards the sensing target direction while a null to the direction of the communication user to alleviate interference.

\begin{figure*}[t]
	\centering
	\subfigure[LoS dominant area]{\includegraphics[width = 0.49\linewidth]{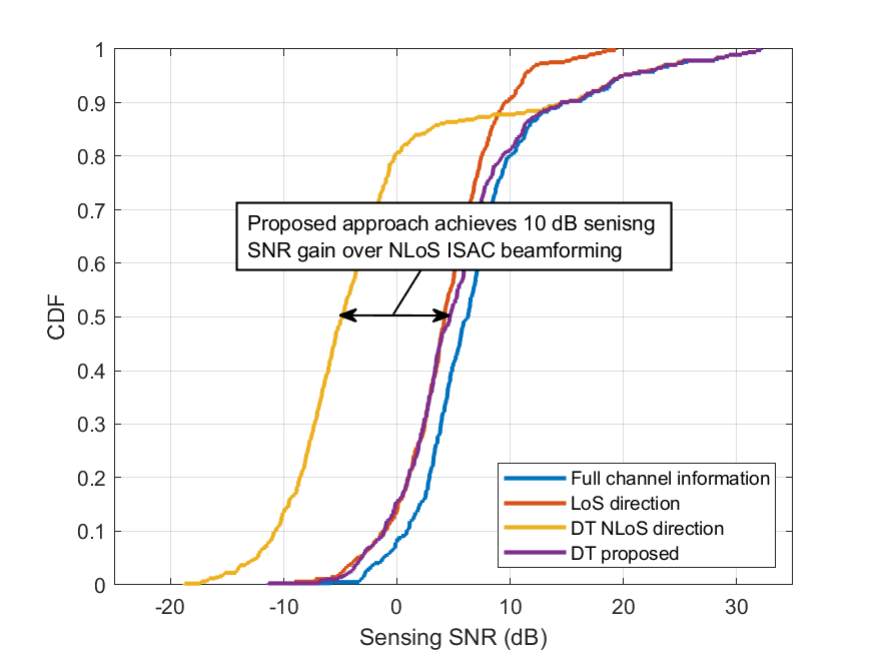}	\label{fig:los}}
	\subfigure[NLoS dominant area]{\includegraphics[width = 0.49\linewidth]{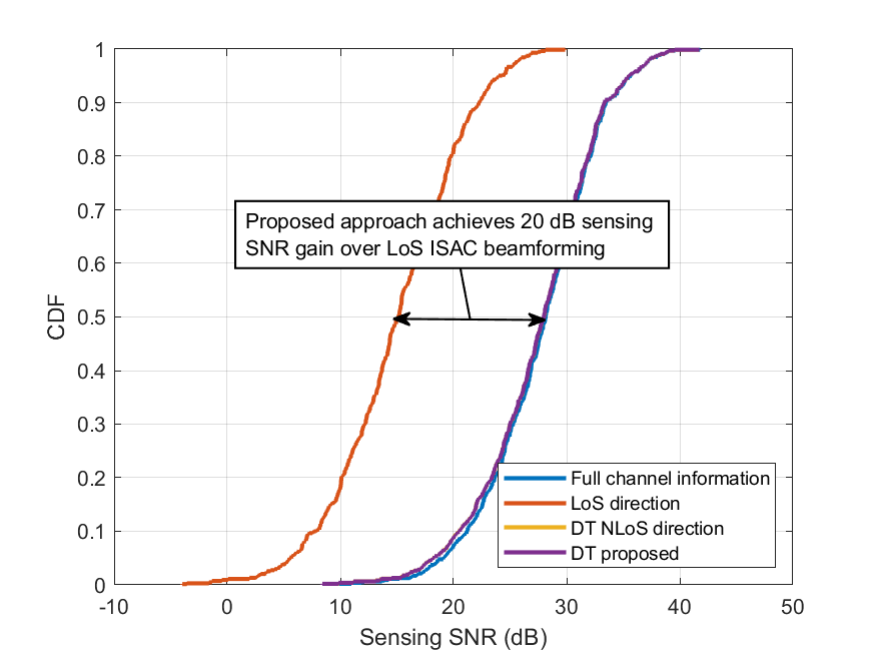}	\label{fig:nlos}}
	\caption{This figure presents the CDF of the sensing SNR with the minimum communication SINR of 10 dB. The proposed digital twin-aided approach achieves near-optimal performance compared to the genie-aided approach, which has full channel knowledge in both the LoS dominant and NLoS dominant areas.}
	\label{fig:sensing_snr_indoor5_all}
\end{figure*}
The scenario in \figref{fig:scenario} can be roughly divided into two areas. In the NLoS dominant area, the glass wall works as a dominant reflector, resulting in a path that reaches the sensing target within one reflection. In this case, the sensing channels are dominated by the NLoS path. For the LoS dominant area, the single-bounce reflection path does not exist, and the sensing channel is dominated by the LoS path that penetrates through the wall. Note that the gain of the single-bounce NLoS path is stronger than the LoS path that penetrates through the concrete wall blockage. Therefore, the sensing channel power in the NLoS dominant area is larger than that in the LoS area.

In \figref{fig:sensing_snr_indoor5_all}, we set the minimum communication SINR to 10 dB, conduct $10,000$ repeated experiments, and compare the empirical cumulative distribution function (CDF) of four approaches. (i) The ``full channel information" is introduced in \sref{sec:full}. (ii) The ``LoS direction" uses the LoS direction information, which is introduced in \sref{sec:los}. (iii) The ``DT NLoS" considers the glass wall as the main reflector and calculates the NLoS direction using the digital twin. This approach then designs the ISAC beamforming by substituting the NLoS direction into \eqref{eq:qt_los}. (iv) The ``DT proposed" uses the dominant partial path direction described in \sref{sec:nlos_approach}.

It can be seen that the ISAC beamforming designed using the full sensing channel information achieves the highest performance in both two areas. It is worth noting that, however, this approach is not practical as the sensing channel is difficult to obtain in real-world applications. Therefore, the performance of the full channel information approach can be considered an upper bound. For the ISAC approach using the LoS direction, it can be seen that the SNR performance is close to that of the full channel information approach in the LoS-dominant area. However, when the target is in the NLoS-dominant area, the LoS approach does not utilize the strong NLoS path. Therefore, the performance gap between the LoS and full channel information approaches is large ($\sim$ $10$ dB). The digital twin-aided NLoS approach achieves near-optimal sensing SNR performance in the NLoS-dominant area. However, its sensing SNR is 10 dB less than the full channel information approach in the LoS-dominant area. Leveraging digital twins, the proposed approach designs the ISAC beamforming according to the most dominant direction in the partial sensing channel. \textbf{The proposed approach achieves near-optimal performance compared with the upper bound using full sensing channel information in both the LoS and NLoS areas}.

\section{Conclusion}
This paper investigates a novel direction that uses digital twins to aid the joint communication and sensing beamforming for MIMO ISAC systems. By applying ray tracing on the EM 3D model of the communication environment, the digital twin can be leveraged to infer the direction and power gain of the propagation paths of the wireless sensing channel. This channel information is then used to guide the joint beamforming design, which maximizes the sensing SNR while satisfying a minimum SINR requirement for the communication link. The proposed digital twin-aided ISAC joint beamforming is evaluated on realistic wireless data generated from high-fidelity ray tracing. Results show that the proposed beamforming approach achieves near-optimal sensing performance while meeting the communication SNR requirement.
\bibliographystyle{IEEEtran}

\begin{thebibliography}{10}
	\providecommand{\url}[1]{#1}
	\csname url@samestyle\endcsname
	\providecommand{\newblock}{\relax}
	\providecommand{\bibinfo}[2]{#2}
	\providecommand{\BIBentrySTDinterwordspacing}{\spaceskip=0pt\relax}
	\providecommand{\BIBentryALTinterwordstretchfactor}{4}
	\providecommand{\BIBentryALTinterwordspacing}{\spaceskip=\fontdimen2\font plus
		\BIBentryALTinterwordstretchfactor\fontdimen3\font minus
		\fontdimen4\font\relax}
	\providecommand{\BIBforeignlanguage}[2]{{%
			\expandafter\ifx\csname l@#1\endcsname\relax
			\typeout{** WARNING: IEEEtran.bst: No hyphenation pattern has been}%
			\typeout{** loaded for the language `#1'. Using the pattern for}%
			\typeout{** the default language instead.}%
			\else
			\language=\csname l@#1\endcsname
			\fi
			#2}}
	\providecommand{\BIBdecl}{\relax}
	\BIBdecl
	
	\bibitem{liu2022integrated}
	F.~Liu, Y.~Cui, C.~Masouros, J.~Xu, T.~X. Han, Y.~C. Eldar, and S.~Buzzi,
	``{Integrated sensing and communications: Toward dual-functional wireless
		networks for 6G and beyond},'' \emph{IEEE J. on Sel. Areas in
		Commun.}, vol.~40, no.~6, pp. 1728--1767, 2022.
	
	\bibitem{demirhan2023integrated}
	U.~Demirhan and A.~Alkhateeb, ``{Integrated sensing and communication for 6G:
		Ten key machine learning roles},'' \emph{IEEE Commun. Mag.}, 2023.
	
	\bibitem{liu2020joint}
	X.~Liu, T.~Huang, N.~Shlezinger, Y.~Liu, J.~Zhou, and Y.~C. Eldar, ``{Joint
		transmit beamforming for multiuser MIMO communications and MIMO radar},''
	\emph{IEEE Trans. on Signal Process}, vol.~68, pp. 3929--3944, 2020.
	
	\bibitem{luo2020multibeam}
	Y.~Luo, J.~A. Zhang, X.~Huang, W.~Ni, and J.~Pan, ``Multibeam optimization for
	joint communication and radio sensing using analog antenna arrays,''
	\emph{IEEE Trans. on Veh. Technol.}, vol.~69, no.~10, pp.
	11\,000--11\,013, 2020.
	
	\bibitem{zhao2022joint}
	N.~Zhao, Y.~Wang, Z.~Zhang, Q.~Chang, and Y.~Shen, ``Joint transmit and receive
	beamforming design for integrated sensing and communication,'' \emph{IEEE
		Commun. Letters}, vol.~26, no.~3, pp. 662--666, 2022.
	
	\bibitem{liyanaarachchi2023joint}
	S.~D. Liyanaarachchi, T.~Riihonen, C.~B. Barneto, and M.~Valkama, ``{Joint MIMO
		communications and sensing with hybrid beamforming architecture and OFDM
		waveform optimization},'' \emph{IEEE Trans. on Wireless
		Commun.}, 2023.
	
	\bibitem{demirhan2023cell}
	U.~Demirhan and A.~Alkhateeb, ``{Cell-free joint sensing and communication
		MIMO: A max-min fair beamforming approach},'' in \emph{2023 57th Asilomar
		Conference on Signals, Systems, and Computers}.\hskip 1em plus 0.5em minus
	0.4em\relax IEEE, 2023, pp. 381--386.
	
	\bibitem{hua2023optimal}
	H.~Hua, J.~Xu, and T.~X. Han, ``Optimal transmit beamforming for integrated
	sensing and communication,'' \emph{IEEE Trans. on Veh.
		Technol.}, 2023.
	
	\bibitem{alkhateeb2023real}
	A.~Alkhateeb, S.~Jiang, and G.~Charan, ``{Real-time digital twins: Vision and
		research directions for 6G and beyond},'' \emph{IEEE Commun.  Mag.}, 2023.
	
	\bibitem{jiang2024learnable}
	S.~Jiang, Q.~Qu, X.~Pan, A.~Agrawal, R.~Newcombe, and A.~Alkhateeb,
	``{Learnable wireless digital twins: Reconstructing electromagnetic field
		with neural representations},'' \emph{arXiv preprint arXiv:2409.02564}, 2024.
	
	\bibitem{alikhani2024large}
	S.~Alikhani, G.~Charan, and A.~Alkhateeb, ``{Large wireless model (LWM): A
		Foundation model for wireless channels},'' \emph{arXiv preprint
		arXiv:2411.08872}, 2024.
	
	\bibitem{luo2010semidefinite}
	Z.-Q. Luo, W.-K. Ma, A.~M.-C. So, Y.~Ye, and S.~Zhang, ``Semidefinite
	relaxation of quadratic optimization problems,'' \emph{IEEE Signal Process.
		Mag.}, vol.~27, no.~3, pp. 20--34, 2010.
	
\end{thebibliography}

\end{document}